\documentclass[]{aastex}
\usepackage{emulateapj5}

\usepackage{graphics}

\newcommand{\mycomment}[1]{}
\newcommand{\UM}{UM\,425}
\newcommand{\UMA}{UM\,425A}
\newcommand{\UMB}{UM\,425B}
\newcommand{\degs}{\ifmmode ^{\circ}\else$^{\circ}$\fi}
\newcommand{\lopt}{\ifmmode L_{2500} \else $~L_{2500}$\fi}
\newcommand{\loglopt}{\ifmmode{\rm log}~L_{2500} \else log$~L_{2500}$\fi}
\newcommand{\logz}{\ifmmode{\rm log}~z \else log$~z$\fi}
\newcommand{\ew}{\ifmmode{W_{\lambda}} \else $W_{\lamzpcbda}$\fi}
\newcommand{\ax}{\ifmmode{\alpha_x} \else $\alpha_x$\fi} 
\newcommand{\aox}{\ifmmode{\alpha_{\rm ox}} \else $\alpha_{\rm ox}$\fi} 
\newcommand{\fcgs}{\ifmmode erg~cm^{-2}~s^{-1[B}\else erg~cm$^{-2}$~s$^{-1}$\fi}
\newcommand{\lnucgs}{\ifmmode erg~s^{-1}~Hz^{-1}\else erg~s$^{-1}$~Hz$^{-1}$\fi}
\newcommand{\lcgs}{\ifmmode erg~~s^{-1}\else erg~s$^{-1}$\fi}
\newcommand{\kms}{\ifmmode~{\rm km~s}^{-1}\else ~km~s$^{-1}~$\fi}
\newcommand{\mone}{\ifmmode ^{-1}\else$^{-1}$\fi}
\newcommand{\mtwo}{\ifmmode ^{-2}\else$^{-2}$\fi}
\newcommand{\msun}{\ifmmode {M_{\odot}}\else${M_{\odot}}$\fi}
\newcommand{\lapprox }{{\lower0.8ex\hbox{$\buildrel <\over\sim$}}}
\newcommand{\gapprox }{{\lower0.8ex\hbox{$\buildrel >\over\sim$}}}
\newcommand{\nh}{\ifmmode{\rm N_{H}} \else N$_{H}$\fi}
\newcommand{\nhgal}{\ifmmode{ N_{H}^{Gal}} \else N$_{H}^{Gal}$\fi}
\newcommand{\nhintr}{\ifmmode{ N_{H}^{intr}} \else N$_{H}^{intr}$\fi}
\newcommand{\nhtot}{\ifmmode{ N_{H}^{tot}} \else N$_{H}^{tot}$\fi}
\newcommand{\atoms}{\ifmmode{\rm ~atoms~cm^{-2}} \else ~atoms cm$^{-2}$\fi}
\newcommand{\cmsq}{\ifmmode{\rm ~cm^{-2}} \else cm$^{-2}$\fi}
\def\Chandra     {{\em Chandra}}

\slugcomment{To appear in Ap.J.}

\shorttitle{Chandra Observation of UM425}
\shortauthors{Aldcroft et al.}

\begin{document}


\title{Lens or Binary? Chandra Observations of the Wide Separation
Broad Absorption Line Quasar Pair UM\,425} 


\author{Thomas L. Aldcroft}
\and
\author{Paul J. Green}
\vskip0.5cm
\affil{Harvard-Smithsonian Center for Astrophysics, 60 Garden Street, Cambridge, MA 02138}
\email{taldcroft@cfa.harvard.edu}



\begin{abstract}
We have obtained a 110 ksec \Chandra\ ACIS-S exposure of UM425, a pair
of QSOs at $z=1.47$ separated by 6.5\arcsec, which show remarkably
similar emission and broad absorption line (BAL) profiles in the
optical/UV.  Our 5000 count X-ray spectrum of UM425A (the brighter
component) is well-fit with a power law (photon spectral index
$\Gamma=2.0$) partially covered by a hydrogen column of $3.8\times
10^{22}$~cm$^{-2}$.  The underlying power-law slope for this object
and for other recent samples of BALQSOs is typical of radio-quiet
quasars, lending credence to the hypothesis that BALs exist in every
quasar.  Assuming the same $\Gamma$ for the much fainter image of
UM425B, we detect an obscuring column 5 times larger. We search for
evidence of an appropriately large lensing mass in our \Chandra\ image
and find weak diffuse emission near the quasar pair, with an X-ray
flux typical of a group of galaxies at redshift $z \sim 0.6$.  From
our analysis of archival HST WFPC2 and NICMOS images, we find no
evidence for a luminous lensing galaxy, but note a 3-$\sigma$ excess
of galaxies in the UM425 field with plausible magnitudes for a $z=0.6$
galaxy group.  However, the associated X-ray emission does not imply
sufficient mass to produce the observed image splitting.  The lens
scenario thus requires a dark (high $M/L$ ratio) lens, or a fortuitous
configuration of masses along the line of sight.  UM425 may instead be a
close binary pair of BALQSOs, which would boost arguments that
interactions and mergers increase nuclear activity and outflows.
\end{abstract}




\section{Introduction}

An important question in current quasar research is whether
powerful mass outflows persist in all quasars throughout their
active lifetimes.  Perhaps only some quasars host these outflows,
which may be characteristic of an early phase of high activity
and accretion rates, possibly triggered by galaxy interactions
and mergers.   

Mass outflows can be studied in detail through intrinsic quasar
absorption lines that hold great promise for revealing the conditions
near the supermassive black holes.  The richest and most extreme
absorption lines are found in quasars with broad absorption lines
(BALs). About 10 - 15\% of optically-selected QSOs have restframe
ultraviolet spectra showing these BALs - deep absorption troughs displaced
blueward from the corresponding emission lines in the high ionization
transitions of C\,IV, Si\,IV, N\,V, and O\,VI (hiBALs hereafter).
About 10\% of BALQSOs also show broad absorption in lower ionization
lines of Mg\,II or Fe\,II (loBALs). BALQSOs in general have higher
optical/UV polarization than non-BAL QSOs, but the loBAL subsample
tends to have particularly high polarization \citep{schmidt99}
along with signs of reddening by dust \citep{sprayberry92, egami96}.
Large samples of BALQSOs from the Sloan Digital Sky Survey (SDSS)
show BAL fractions of about 1/3 at the redshift of peak selection
efficiency, and increasing reddening in the sequence non-BAL, hiBAL,
loBAL has now been verified \citep{reichard03}.  
All the BALs are attributed to material along our line
of sight flowing outward from the nucleus with velocities of up to
$\sim 50,000$\kms.  Emission line flux is not observed at comparable
velocity widths, so if flux is scattered from the BAL material, 
its must cover $<20\%$ of the BAL region \citep{hamann93}. Together
with the similar fraction of QSOs  showing BALs, this suggests that most or
possibly {\em all} QSOs contain BAL-type outflows, which are only seen
along sightlines traversing the BAL  clouds.  In this orientation
scenario, BALQSOs provide a unique probe of conditions near the
nucleus of most QSOs.  If so, we expect that their {\em intrinsic}
X-ray emission should be consistent with those of normal QSOs.  Recent
studies with \Chandra\ \citep{green01, gallagher02a} support this
orientation interpretation; once absorption columns of $\nhintr \geq 
10^{22}\atoms$ are accounted for, the underlying X-ray power-law
continuua appear to have typical slopes and normalizations.

An important alternative interpretation is that BALQSOs are instead
adolescent quasars in an outburst or transition phase, expelling a
cocoon of circumnuclear gas and dust while evolving from active high
$L/L_{\rm Edd}$ (high Eddington fraction) QSOs toward normal QSOs
\citep{hazard84, gregg02}. 
Links between low-ionization BALQSOs and IR-luminous mergers
\citep{canalizo01, fabian99}, and similarities between BALQSOs and and
narrow line Seyfert~1 galaxies \citep{mathur00, brandt00} support this
scenario.  Furthermore, since the outflows are thought to contain high
metallicity gas \citep{hamann99, arav01}, BALQSOs may be relevant
to studies of the formation and early (high redshift) evolution of
galactic nuclei.  It is intriguing that approximately half of the $z\geq 5$
QSOs found so far in the Sloan Digital Sky Survey show BALs
\citep{zheng00, fan03}.

BALQSOs that occur in multiples hold particular interest.  Those that
are gravitationally lensed may be magnified, and possibly microlensed,
providing an opportunity for study of the quasars' intrinsic absorbers
along slightly different sightlines \citep[e.g.][]{lewis02}.  Several
lensed BALQSOs are known, but differential absorption studies are
difficult, especially in X-rays, due to their close (typically
$\sim1\arcsec$) spacing.  Several multiple BALQSOs with wider
separations ($\geq 3\arcsec$) lack clear lens candidates.  If they are
lensed, they provide a wider binocular view of the intrinsic
absorbers.  If not, 
then as binary pairs their ($\sim$ tens of kpc) linear separations
allow practical study of the dynamical interactions proposed to spur
high activity and mass outflow phases.

\UM\ is a pair of BALQSOs at redshift $z=1.465$ discovered by
\citet{meylan89} in a search for lenses using selection of bright,
high redshift (presumably magnified) quasars. Separated by 6.5\arcsec,
the 2 brightest images have nearly identical optical/UV spectra and
close velocities: $\Delta v_{A-B}=200\pm100$\kms\ from
\citet{meylan89}; and $\Delta v_{A-B}=630\pm130$\kms\ from
\citet{michalitsianos97}.  Deep imaging to $R \sim 24$ reveals no
deflector, arcs, or arclets \citep{courbin95}, whereas a massive lens
should be present to cause the large separation.  We observed \UM\
with \Chandra\  on 2000-Apr-07 as part of an X-ray snapshot survey of
BALQSOs.  In that survey, we found that as a class BALQSOs are heavily
obscured, but otherwise normal radio-quiet quasars \citep{green01}.
Citing the relatively high count rate (0.04 cts/sec) for a BALQSO and
the unique binary/lens nature of the of \UM\ system, we proposed for
120 ksec of followup imaging spectroscopy in \Chandra\  cycle 3.  In this
paper we describe the results of this observation, including spectral,
timing, and image analysis of the data.  We also marshall evidence
from previously unpublished archival HST data taken with STIS, WFPC2,
and NICMOS.

\section{Observation and data reduction}

\begin{figure*}
\centerline{\resizebox{4.0in}{!}{\includegraphics{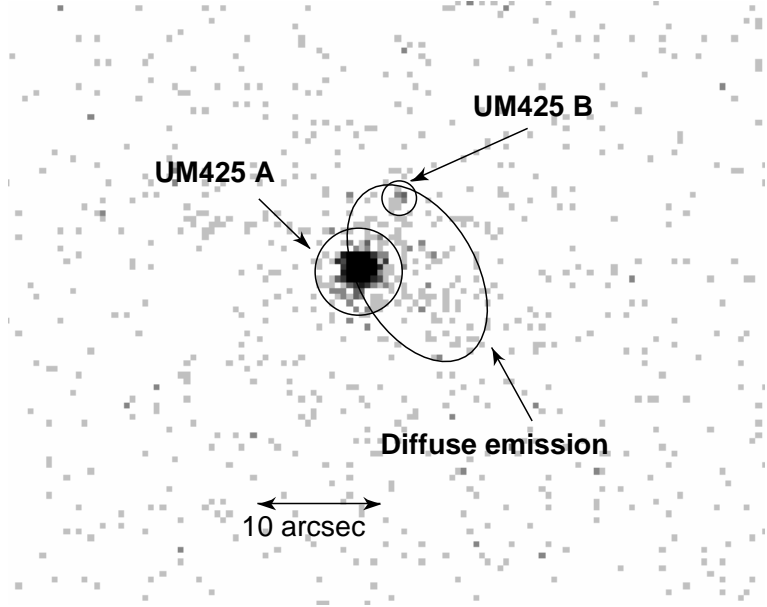}}}
\caption{Soft-band image (0.3-3.0~keV) of 110 ksec ACIS-S exposure of \UM.
North is up, East to the left.  
In addition to the clearly detected A and B components of \UM, there is
evidence for a diffuse component centered about 5~arcsec west of \UMA. 
The elliptical region has an excess of $51\pm12$ counts over the
expected counts due to \UMA\ and the background.}
\label{fig:image}
\end{figure*}

\UM\ was observed for 110~ksec on 2001-Dec-13 (ObsId 3013) at the
nominal aimpoint of ACIS-S3 using 3.14 second full-frame readouts in
timed very faint mode.  A net of 4927 counts between 0.3~keV and 8.0~keV
were detected for \UMA, and 28.3 counts for \UMB.  The soft-band image
(0.3-3.0~keV) of \UM\ is shown in Figure~\ref{fig:image}, with
0.5\arcsec\ spatial binning corresponding to the ACIS CCD pixel size.
The pixel intensities have been scaled logarithmically.  The X-ray
celestial coordinates match the optical counterpart coordinates to
within $\sim 1\arcsec$, so there is no ambiguity about identification.
Furthermore the relative separation of the X-ray A and B components
matches the optical separation to within 0.3\arcsec.  In addition to
the two point sources associated with \UM, we also see evidence for
faint diffuse emission which could be associated with a foreground
cluster or group of galaxies.  This emission is significant at the
4.2-$\sigma$ level, and is discussed in Section~\ref{sec:diffuse}.
Finally, there is a noticable extended linear feature approximately
25\arcsec\ N-NW of \UMA\ that coincides with a bright foreground galaxy
in the field.

The \Chandra\  X-ray observation data was produced by the CXCDS automatic
processing pipeline, version 6.4.0.  To take advantage of
subsequent improvements in the ACIS response and gain calibration files, we
used the CIAO tool {\tt acis\_process\_events} to update the event file.  At
the same time, pixel randomization was turned off in order to allow spatial
analysis at the finest level of detail.  CIAO version 2.3 and CALDB version
2.10 were used in all data analysis and processing tasks.

The spectral data reduction followed the standard CIAO 2.3 thread to 
extract an ACIS spectrum for \UMA : (1) Extract source events
within a 4.0\arcsec\ radius, using a background annulus spanning 8 to
30\arcsec; (2) Create the aspect histogram file; (3) Create the RMF
and ARF files appropriate to the time-dependent source position on
chip.  The ARF was corrected for the ACIS time-dependent quantum efficiency
degradation using the \texttt{corrarf} 
program\footnote{http://cxc.harvard.edu/cal/Acis/Cal\_prods/qeDeg/index.html}.
  In addition to the standard thread, the event data were
filtered on energy to use the range 0.3-8.0 keV and they were grouped
to an average of 30 counts per bin.  

It should be be noted that the region used to extract the spectrum for
component A includes some of the extended emission discussed in
Section~\ref{sec:diffuse}. However, this component contributes fewer than $\sim
10$ counts, and therefore does not noticably affect the spectral fitting.


\section{Spectral Modeling}
We carried out spectral modeling of sources in the \UM\ field using
\textit{Sherpa}, a generalized modeling and fitting environment within CIAO.
We fit using $\chi^2$ data-variance statistics with Marquardt-Levenberg
optimization.  The energy range used for fitting was 0.5-8~keV, which avoids
the region below 0.5~keV that is not well calibrated.

\subsection{\UMA}

\begin{figure*}
\centerline{
\resizebox{4.5in}{!}{\includegraphics{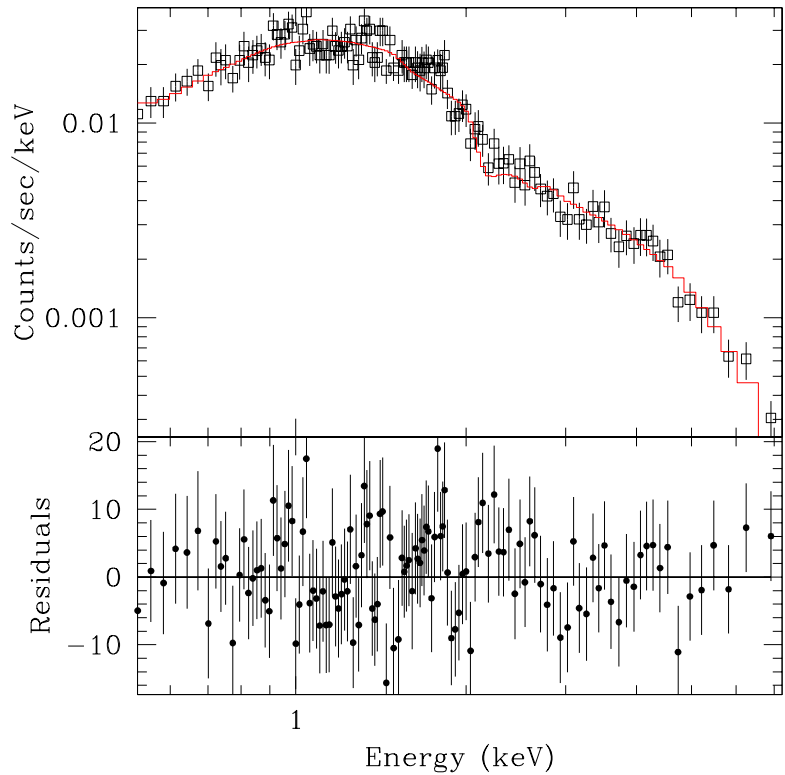}}
}
\caption{Spectral fit of \UMA\ with a powerlaw ($\Gamma=2.0$), a partially
covering neutral absorber ($N_H = 3.8\times10^{22}~$cm$^{-2}$ and covering
fraction $f_{PC}=0.73$) at the quasar redshift, and Galactic neutral absorption
($N_H = 4.1\times10^{20}~$cm$^{-2}$).}
\label{fig:spec_fit}
\end{figure*}

\begin{figure*}
\centerline{
\resizebox{3.0in}{!}{\includegraphics{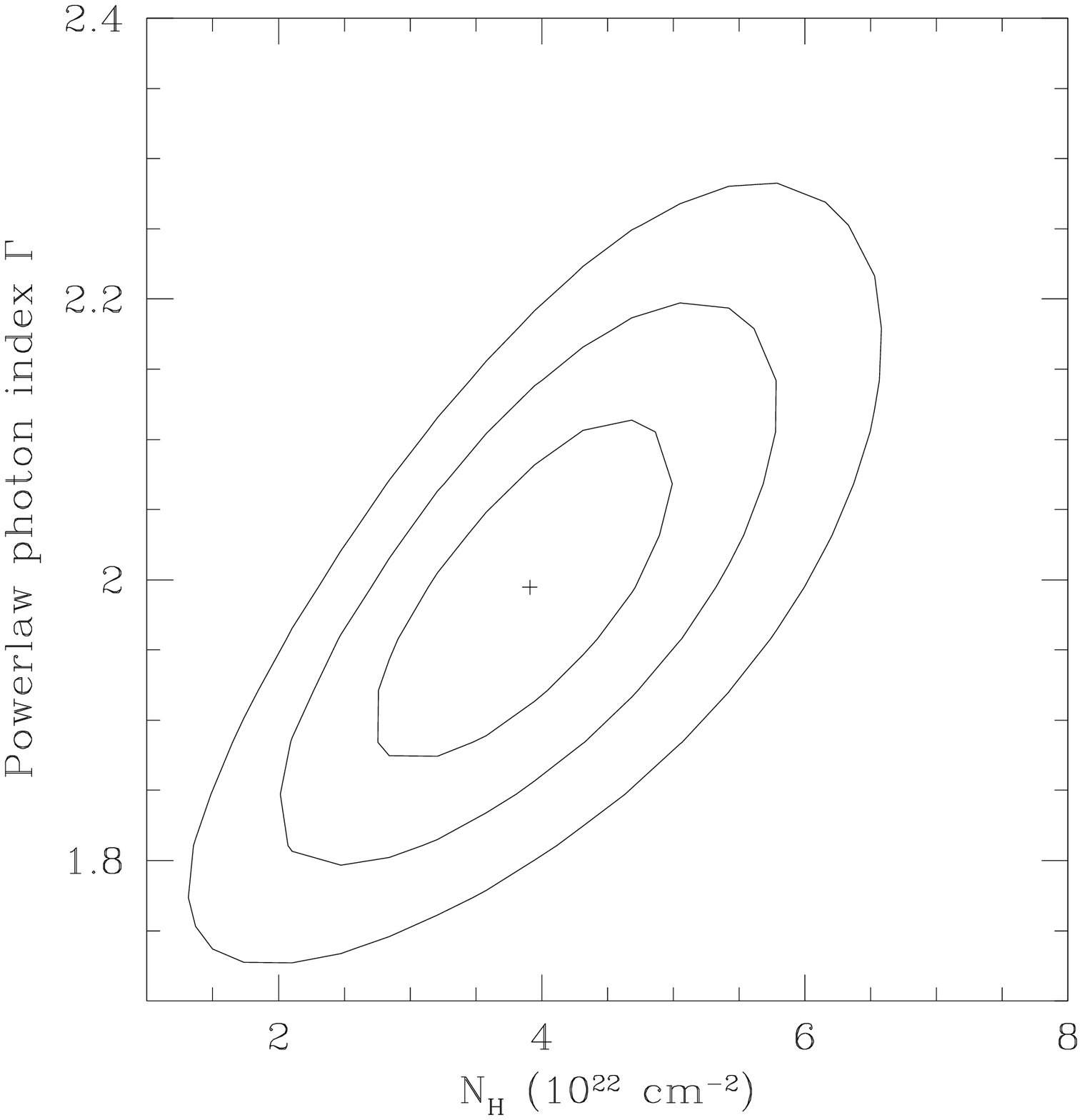}}
\resizebox{3.0in}{!}{\includegraphics{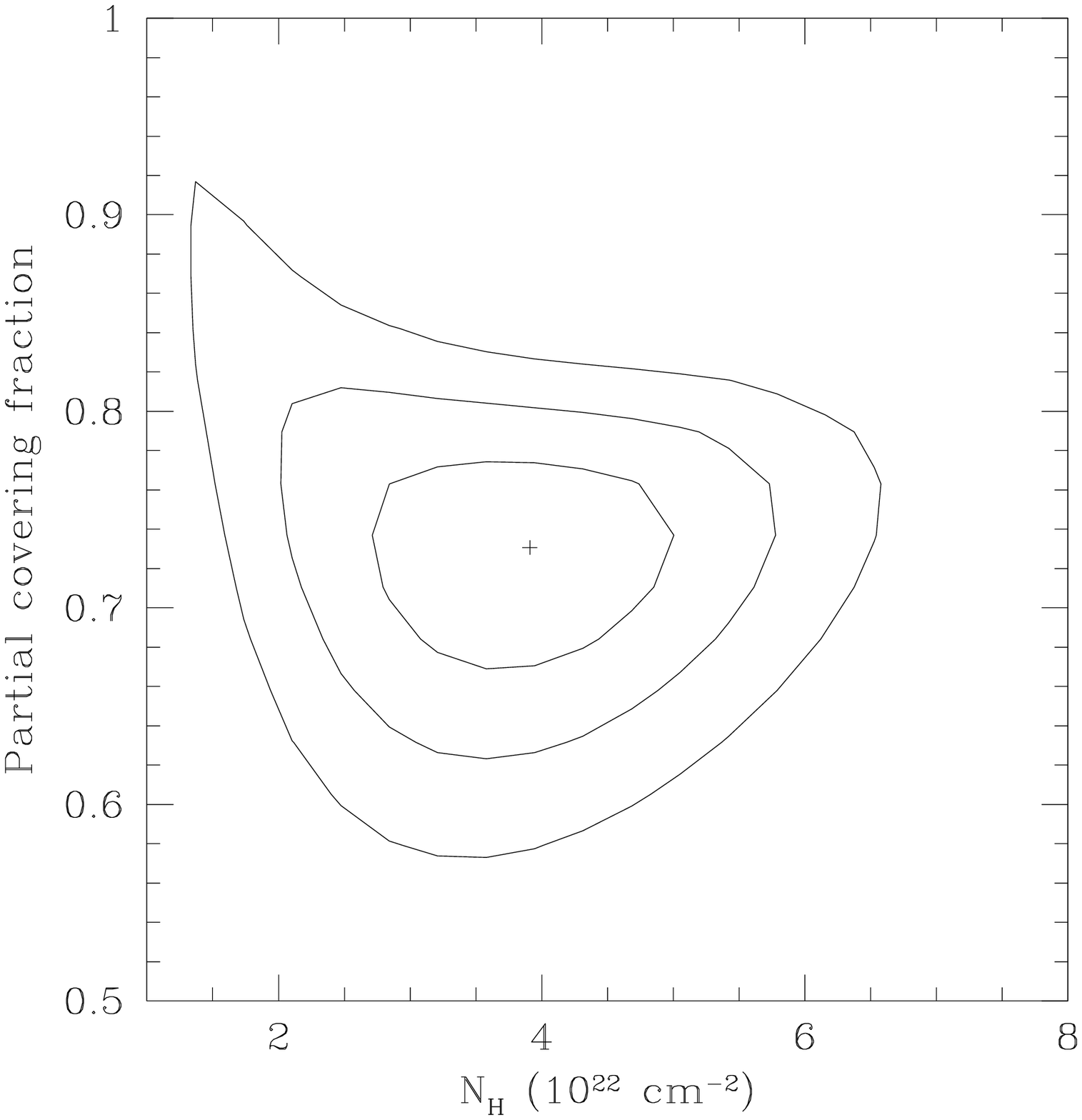}}
}
\caption{Confidence intervals (1,2,3-$\sigma$) 
for redshifted intrinsic absorption versus power-law spectral index $\Gamma$
(left panel) and for redshifted intrinsic absorption versus covering fraction
(right panel) in our fits to the X-ray spectraum of \UMA.}
\label{fig:contour}
\end{figure*}

The results of spectral modeling of the \UMA\ spectrum are given in
Table~\ref{tab:fit}.  After taking account of the Galactic column of $\nhgal=
4.1\times10^{20}$\atoms, the X-ray spectrum of \UM\ is well-fit with either a
partially-covered neutral absorber or a highly-ionized warm absorber.  For
neutral absorption, partial covering is required. The best fit neutral
absorption model and residuals are shown in Figure~\ref{fig:spec_fit}.  The
underlying spectrum of \UM\ is consistent with the spectrum of a normal
radio-quiet quasar at $z \sim 1.5$ absorbed by an intrinsic column of
$3.8\times 10^{22}$~cm$^{-2}$.  The power-law index $\Gamma = 2.0$
for \UM\ agrees with the value of $\sim 1.9$ seen with ASCA for RQQs at
redshifts $1.5 < z < 2.5$ \citep{reeves00}.  Confidence contours for powerlaw
slope $\Gamma$ and partial covering fraction are shown plotted against
intrinsic absorbing column $N_{H,z}$ in Figure~\ref{fig:contour}.

\begin{table*}[t]
\small
\caption{\sc X-ray Spectral fit parameters for \UMA}
\label{tab:fit}
\begin{tabular}[t]{lcrrcccc}
\hline \hline \\
\hfil Model \hfil  &  $\Gamma$  &  Amplitude   &  $N_{H,z} $ & Other & Flux & $\chi^2$ (DOF) \\
       &  (a)    &  (b) \hspace*{3ex}                  &  (c)\hspace*{1ex}  &(d) &(e)& \vspace{0.5ex}  \\
\hline \vspace{-2.5ex} \\
\hline  &\vspace{-2.0ex} \\
Gal $N_H$ (fixed)     & $1.43\pm0.04 $ & $5.3\pm0.2$ &  ...        & ...                  &$3.7\pm0.1$& 430.5(122) \vspace{0.5ex} \\
$N_H (z=1.465)$       & $1.78\pm0.08 $ & $7.9\pm0.6$ & $1.1\pm0.2$ & ...                  &$3.4\pm0.1$& 145.3(121) \vspace{0.5ex} \\
Part. Cov. $N_H (z=1.465)$  			     
                      & $1.99\pm0.13 $ &$10.4\pm1.6$ & $3.8\pm1.2$ & $f_{PC}=0.73\pm0.06$ &$3.4\pm0.1$& 122.2(120) \vspace{0.5ex} \\
Warm absorber (CLOUDY)& $2.00\pm0.06 $ &$11.6\pm2.0$ &$10.0\pm1.5$ & $U=1.76\pm0.04$      &$3.4\pm0.1$& 126.3(120) \vspace{0.5ex} \\
\hline \vspace{-1.5ex} \\ 
\multicolumn{7}{c}{
\parbox{7in}{
\small
{\sc Notes:}  Uncertainties are 90\% 
confidence limits. 
(a) Power law photon index.  (b) Power law normalization in units of $10^{-5}$\,photons\,cm$^{-2}$\,s$^{-1}$\,keV$^{-1}$ at 1\,keV
(c) Absorbing column in units $10^{22}$~cm$^{-2}$ at quasar redshift. 
(d) $f_{PC}$ is the partial covering fraction, $U$ is the $\log_{10}$ of the 
dimensionless CLOUDY ionization parameter
(e) Model flux (0.3-8~keV) in units $10^{-13}$~ergs\,cm$^{-2}$\,s$^{-1}$
}}
\end{tabular}
\end{table*}

Given the observed count rate of 0.045~cts\,s$^{-1}$ and the frame
read time of 3.1~sec, we estimate (using
PIMMS\footnote{http://asc.harvard.edu/toolkit/pimms.jsp}) a pileup
fraction of approximately 6\%. Using the \verb|jdpileup| model
\citep{davis01}, we verify that this level of pileup does not have a
statistically significant impact on our best-fit spectral model
parameters.

The unabsorbed 0.5-2~keV flux is $1.13\times 10^{-13}$~\fcgs.  
Using that de-absorbed X-ray flux, the optical to X-ray flux
ratio\footnote{$\aox$\, is the slope of a hypothetical power-law from
2500\,\AA\, to 2~keV; $\aox\, = 0.384~{\rm log} \frac{\lopt}{L_{2\rm keV}}$} is
$\alpha_{ox} = 1.6$, which is consistent with values for normal (non-BAL) QSOs
at this redshift \citep{green95}.
A trend of increasing \aox\, with luminosity has been noted
by many authors (e.g., \citealt{avni82, green95}), but
most recently by \citet{vignali03}.  With its apparently large
optical luminosity (log\,$L_{\nu}=32.18$~\lnucgs\, or
log\,$\nu\,L_{\nu}=46.60$ \lcgs, see Figure~\ref{fig:opt_lum}), the
de-absorbed \aox\, for \UMA\, falls well within the rather large
dispersion in this relationship.   

Distinguishing between neutral and ionized absorption is not possible
at this time due to systematic uncertainties in the ACIS response
calibration in the critical region between 0.2-0.5~keV.  An
additional obstacle is the unfortunate coincidence between the
expected location of O\,{\sc vii} absorption and the instrumental
Carbon edge (284 eV).  

\subsection{Iron lines and edges in a BALQSO spectrum}
\citet{chartas02} found strong relativistic broad absorption lines (at
rest energies 8.1 and 9.9~keV) in the ACIS-S spectrum of the $z$=3.91
BALQSO APM\,08279+5255.  With XMM data for this same source,
\citet{hasinger02} found instead an edge at rest energy 7.7~keV.
In both cases these features were attributed to highly ionized Fe associated
with the BAL outflow.  In \UMA\ we find no evidence for such features, nor do
we see any residuals that are inconsistent with the systematic uncertainties in
the ACIS response calibration.  If the same absorption troughs or edges were
present \UMA\ with a similar strength, we would have detected them at 3 to
5-$\sigma$ confidence.



\subsection{\UMB}
\label{sec:umb}

The observed ratio of broadband (0.3-8~keV) source counts for the A and B
components of \UM\ is 175.  If \UM\ is truly a lensed system, this ratio would
be well above the naively expected value of $\sim 60$ based on the 4.5~mag
difference in R magnitude \citep{courbin95}.  However, this does not account
for the possibility of differing absorbing columns and/or dust-to-gas
ratios along the two sightlines, especially considering that the
X-ray emission region is much smaller than the optical.
In fact, if we consider only the hard band counts (2.5-8~keV), which are
largely unaffected by absorption, the ratio is $70\pm 20$.  

Even though \UMB\ has only 29 counts, we can test for the presence
of differing absorption by fixing its X-ray power-law slope to 
the best-fit value for \UMA.  Using the $\Gamma=2.0$ partial
covering model in Table~\ref{tab:fit} and freezing all parameters
except for $N_{H,z}$ and Amplitude, we find a best fit column
of $N_{H,z}=2.0^{+2.6}_{-1.1}\times10^{23}~$cm$^{-2}$ (90\% confidence).
With the assumption of the same underlying continuum, we find that \UMB\ has a
factor of 5 larger intrinsic absorbing column than \UMA, and that the two
spectra are inconsistent at approximately 3-$\sigma$ confidence.
The implications of this difference are discussed further in
Section~\ref{sec:discussion}. 

If \UM\ is not lensed, then this analysis only applies to the extent that a
power law spectrum with $\Gamma=2.0$ is typical of QSOs at $z \approx 1.5$. 
However, with just 29 counts, we cannot usefully constrain both
$\Gamma$ and $N_{H,z}$. 

\subsection{Hardness ratio}
\label{sec:hardness}
An independent, model-free measure of spectral similarity is afforded
by comparing the hardness ratio $${\rm HR}=\frac{H-S}{H+S}$$ of the
two X-ray images.  We perform photometry in 3 energy bands: soft ($S$;
0.3-2.5keV), hard ($H$; 2.5-8keV), and broad ($B$; 0.3-8keV).  We
extract 4927 $B$ band counts from \UMA, and 28 from B.  Given the low
background count rate in the ($B$ band) image of $5\times
10^{-7}$pixel\mone sec\mone, errors may be considered as strictly
$\sqrt{B}$.  \UMA\ has 3929 $S$ and 998 $H$ counts, so ${\rm HR}_A=
-0.59\pm 0.01$.  Since \UMB\ has 14 counts in both $S$ and $H$ bands
for ${\rm HR}_B=0.0\pm0.2$, it is harder at the $3\sigma$
level.\footnote{The uncertainties were calculated using standard error propogation
assuming gaussian errors.  For the \UMB\ values (14 counts in the $S$ and $H$
bands) the error distribution deviates slightly from gaussian.  However, we
have verified by direct Monte-Carlo simulation that the standard deviation of
HR$_B$ is 0.2, and that the hardness ratios HR$_A$ and HR$_B$ are inconsistent
at approximately 99.7\% (3-$\sigma$) confidence.}
This is consistent with our results from fitting of
a power law spectral model.

\section{Variability}
Temporal variability is a potentially key diagnostic for constraining
the absorber geometry for BALQSOs.  \citet{gallagher02b} discovered
hard-band variability at the 45\% level on a timescale of
20~ksec in the nearby loBAL QSO Mkn~231.  From this they inferred an
absorbing geometry in which only indirect, scattered X-rays from
multiple lines of sight are observed, with a small Compton-thick
absorber blocking the direct X-rays.  

\begin{figure*}
\centerline{
\resizebox{4.5in}{!}{\includegraphics{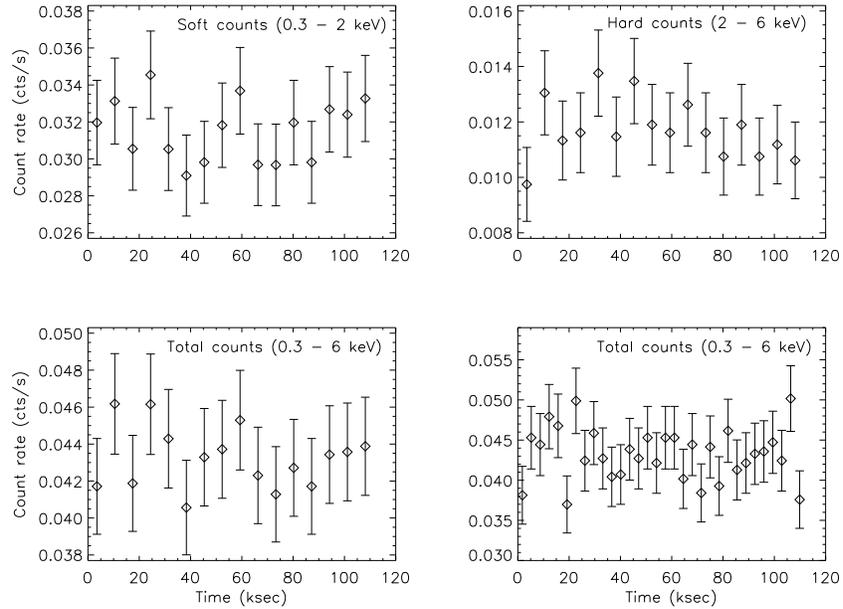}}
}
\caption{Count rate as a function of time for \UMA.  The upper panels
show the rates in the soft and hard bands for 16 equally spaced time bins 
covering the 110~ksec observation. The lower panels show the
broadband count rates, where  the left panel has 16 bins and the right
panel has 32 bins.}
\label{fig:variability}
\end{figure*}

\UMA\ shows no significant short-term variability in either the broad,
soft, or hard bands.  Figure~\ref{fig:variability} shows the count rate as a
function of time.  The upper panels show the soft and hard bands for
16 equally spaced time bins covering the 110~ksec observation, and the
lower panels show the broadband count rates, where the left panel has
16 bins and the right panel has 32 bins.  The error bars are simply
the square root of the number of counts in each bin.  In all cases if
we fit the data with a constant value (no variability) we find
$\chi_{\nu}^2 \la 1$.  Furthermore, over the much longer timescale of
1.7~years spanning the two \Chandra\  observations, 
the count rates in the energy band 1.5-8~keV are consistent to within
1-$\sigma$.  Within this energy band there is negligible change in effective
area due to the ACIS quantum efficiency degradation over 1.7 years, and we can
simply compare count rates.

As another test for variability, we used the Bayesian block
method\footnote{http://astrophysics.arc.nasa.gov/$\sim$jeffrey}$^,$\footnote{S-lang
implementation of the algorithm kindly provided by M. Nowak (CXC/MIT)}
\citep{scargle03,scargle98} to characterize the lightcurve of \UMA. 
This algorithm searches the unbinned event data for statistically significant
changes in the event rate.  We set the detection threshold to 2-$\sigma$
confidence and the code found no rate changes above that significance level
(for broad, soft, and hard band events).


\section{Diffuse emission}
\label{sec:diffuse}

\begin{figure*}
\centerline{
\resizebox{4.0in}{!}{\includegraphics{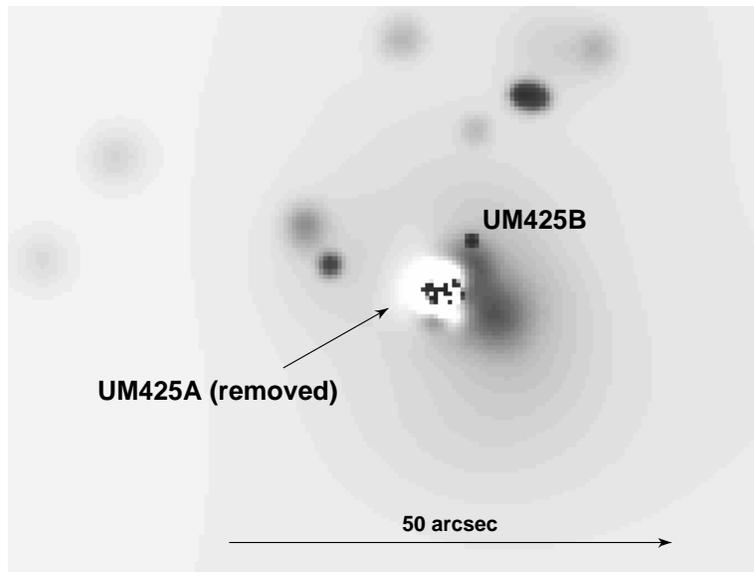}}
}
\caption{Adaptively smoothed soft-band (0.3-3~keV) image of the \UM\ field 
with the central \UMA\ point source subtracted.  North is up, East to
the left.  An elliptical component of 
diffuse emission centered about 5\arcsec\ west of \UMA\ is seen.  The central
15\arcsec\ core of the emission contains $\sim 50$~photons, giving an
integrated count rate of $4.6\times10^{-4}$~cts\,s$^{-1}$. The bright
source $\sim 25$\arcsec\ N-NW of \UMA\ corresponds to a bright
foreground galaxy.}
\label{fig:diffuse_emission}
\end{figure*}

In Figure~\ref{fig:diffuse_emission} we show an adapatively smoothed image of
the soft photons in the UM425 field, where we have subtracted the central
bright point source \UMA.  The image was created in the following manner: We
first filtered the event list to include only photons in the 0.3-3.0~keV
range, then used the CIAO tool \texttt{csmooth} to adaptively smooth with a
minimum significance level of 3-$\sigma$.  We calculated the exposure map,
adaptively smoothed it with the same smoothing scale map, and divided so as to
create a flattened exposure-corrected image.  Next we used our best
fit (partial covering) spectral model as input to 
ChaRT\footnote{http://asc.harvard.edu/chart/} and
MARX\footnote{http://space.mit.edu/CXC/MARX/} to create a simulated PSF at the
position of \UMA.  This PSF image was scaled to the same flux, smoothed with
the original smoothing scale map, aligned with the centroid of \UMA, and
subtracted from the exposure-corrected \UMA\ image. 
Due to the slight pileup (estimated at 6\%) in our observation, the
core flux may slightly underpredict the true incident flux.  We
estimate this could contribute a residual of no more than 11 photons outside
of a 3\arcsec\ radius in the smoothed image of
Figure~\ref{fig:diffuse_emission}.
 Within about 3\arcsec\ of
the core of \UMA\ the residuals are large compared to the faint diffuse flux
(due to a combination of counting noise, pileup, and a slight mismatch in shape), but
outside this radius the PSF subtraction effectively removes the contribution
from \UMA.  Note that within the 3\arcsec\ radius the residuals are no more
than 3-$\sigma$, and the net subtracted counts are consistent with zero.

\begin{figure*}
\centerline{
\plotone{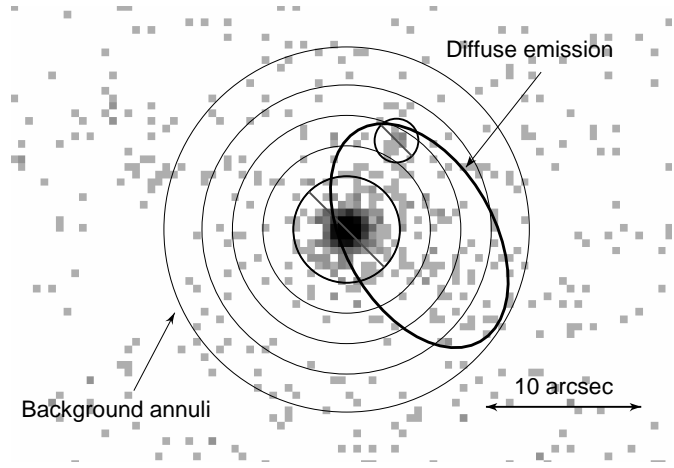}
}
\caption{Elliptical extraction region used to estimate a lower limit on the
diffuse flux.  Source regions containing \UMA\ and \UMB\ are excluded.  For
each annulus, source counts are extracted from within the elliptical region and
background counts are extracted from outside the ellipse.  Net counts from the
four annuli are summed to give a total of $51\pm13$ photons in the 0.3 -
3.0~keV band.  The annular radii are at 3.5, 5.5, 7.5, 9.5, and 12.5\arcsec.
North is up and East is to the left.}
\label{fig:diffuse_counts}
\end{figure*}

Faint diffuse emission is clearly evident in the smoothed images.  This
emission extends at least 20\arcsec\ and is elliptical, centered about 5\arcsec\
to the west of \UMA.  While the extended emission is highly significant,
determining the net diffuse flux requires some care because of the very bright
($\sim$4100 photons between 0.3-3~keV) point source.  Due to small
imperfections in the HRMA, we can expect about 30-60 photons from \UMA\ to be
scattered outside a 10\arcsec\
radius.\footnote{http://cxc.harvard.edu/cal/Hrma/psf/PSF\_wings\_3c273/psf\_wings.html}
Using the PSF-subtracted smoothed image may therefore give unreliable results
when the diffuse component itself has only $\sim 50$ photons.  Instead, we can
estimate a lower limit on the net diffuse flux by considering only the
elliptical core of emission  and
carefully subtracting the background from thin concentric annuli centered on
\UMA\ (see Figure~\ref{fig:diffuse_counts}). This analysis is done on the
filtered event data, and allows us 
to determine the excess without relying on complex PSF models.  Instead we make
use of the axial symmetry of the on-axis PSF.  We find that in the soft band
between 0.3-3.0~keV, there is an excess of $51\pm12$ counts.  In contrast,
between 3.0-8~keV, the extended elliptical region has a net of $-4\pm9$
counts, consistent with zero.  Looking in more detail within the soft band, we
find the counts are split roughly evenly between the energy bands 0.3-0.8~keV
and 0.8-3~keV.

If these 51 photons originate in a Raymond-Smith plasma with a rest
frame temperature $kT = 1.5$~keV, the observed flux $f = 1.5\pm0.4 \times
10^{-15}$~erg\,s$^{-1}$\,cm$^{-2}$ (in either a 0.3-3.0~keV 
or 0.1-2.4~keV band).  This model is fully consistent with the
broadband energy distribution of counts.
At a redshift of 0.6 this model and flux correspond to an X-ray
luminosity of $L_X$(0.1-2.4~keV)$ = 2.7\times
10^{42}$~erg\,s$^{-1}$.\footnote{We use a $H_0 =
70$~km\,s$^{-1}$\,Mpc$^{-1}$, $\Omega_{\Lambda} = 0.7$, and
$\Omega_{M} = 0.3$ cosmology throughout.}  The luminosity and assumed
plasma temperature we derive is consistent with values for groups of
galaxies measured by \citet{mulchaey98} using ROSAT, but a factor of
$\sim 5$ below the expected value for a cluster of galaxies massive
enough to cause the observed image separation (discussed further in
\S~\ref{sec:discussion}).  

The intriguing possibility remains\footnote{Thanks to the anonymous
referee for suggesting we add this discussion.} that the diffuse emission
originates from a cluster at the redshift of \UM, as discussed
by \citet{mathur03}.  This would make this candidate cluster
among the most distant known, second only to the
vicinity of 3C\,294 at $z = 1.786$ \citep{fabian01}, where a
small excess of emission is detected, associated with the southern
radio hotspots.  For the diffuse emission near \UM, further
observations and analysis are needed to determine the redshift of the
candidate cluster, which if coincident with \UM\ has luminosity
$L_X$(0.1-2.4~keV)$ = 2.8\times 10^{43}$~erg\,s$^{-1}$.
Similar diffuse emission around bright quasars is detected around serendipitous
\Chandra\  quasars in \citet{green03}.


\section{UV spectra}

\UMA\ and \UMB\ were observed using HST/STIS (PI T. Gull) with the
G230LB grating for a total of 11800~sec on 1998-Mar-10.  The
observations were done at a roll 
angle of 197.3$^{\circ}$ so that both objects would fall in the $52\times
2$\arcsec\ slit.  The results from this observation have not been previously
published, so we retrieved the data from the HST archive and extracted spectra
for the A and B components of \UM.  The first step in our data reduction was
using the STSDAS {\tt ocrreject} task to reject cosmic rays and combine the
five CCD exposures into a single image.  The final spectral extraction was then
done with a custom IDL code which filtered out the bad pixels which remained
after running {\tt ocrreject}.

\begin{figure*}
\centerline{
\resizebox{7in}{!}{\includegraphics{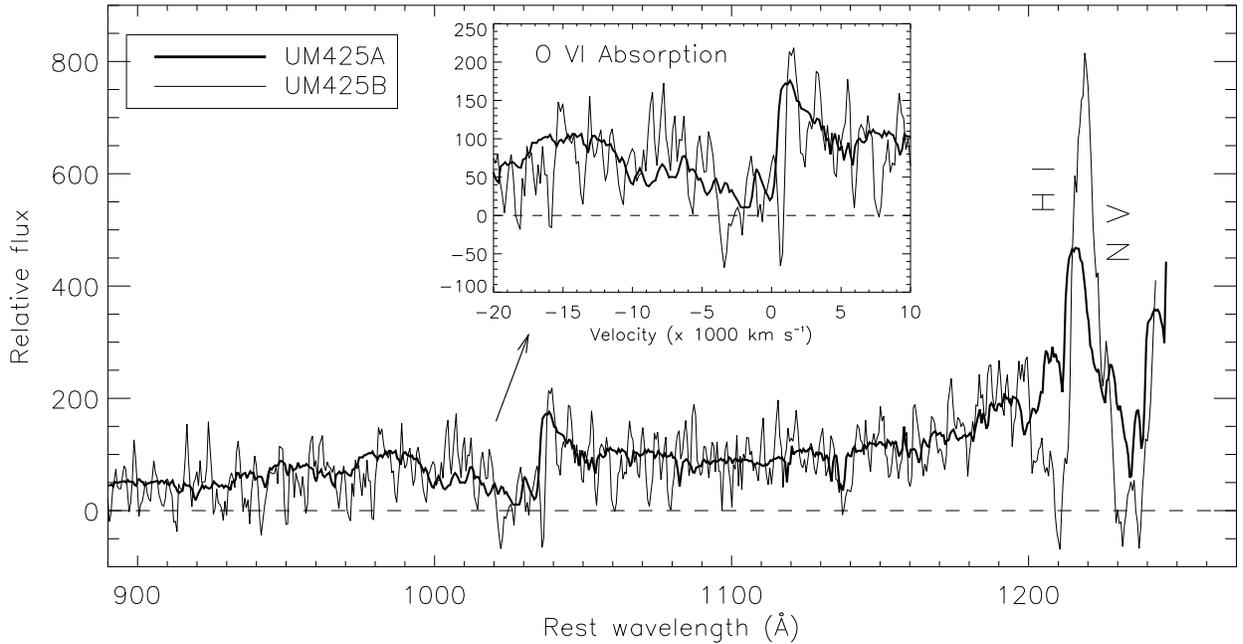}}
}
\caption{Simultaneous HST STIS spectra of \UMA\ and \UMB,
scaled to a common mean value between 1070\,\AA - 1180\,\AA.  \UMA\ is the
bold curve.  The absorption troughs due to N\,V, Ly $\alpha$
are marked, with the inset showing the O\,VI absorption on a velocity scale
relative to the QSO redshift. }
\label{fig:stis}
\end{figure*}

In Figure~\ref{fig:stis} we show an overlay of the STIS spectra for \UMA\
(heavy line) and \UMB\ (light line).  The spectra have been normalized to have
the same mean flux in the rest wavelength range 1070 - 1180\,\AA, which is
devoid of strong absorption and emission features.  In this range we find a
flux ratio A/B of 102, consistent with the ratio of $\sim 100$ seen by
\citet{meylan89} in the 1500 - 2400\,\AA\ rest wavelength range.  In the plot
the wavelength of the \UMB\ spectrum has been shifted by 8.8\,\AA\ (observed
frame) so that the difference in redshift (based on the peak of Ly-$\alpha$) is
equal to the value of $z_A - z_B = 0.006$ found by \citet{michalitsianos97}.
This offset in the relative wavelength calibration for the B component is not
excessive given the 2\arcsec\ wide slit which was used.  For this instrument
configuration, an 8.8\,\AA\ wavelength offset would result if the B component
were off-center in the slit by 0.325\arcsec.

The plot shows that both \UMA\ and \UMB\ have broad absorption lines due to
O\,VI, H\,I Ly-$\alpha$, and N\,V.  The absorption for O\,VI appears to extend
to about 13000~km\,s$^{-1}$ in both \UMA\ and \UMB, although with the limited
S/N in \UMB\ the presence of absorption beyond 5000~km\,s$^{-1}$ is less
certain.  Near the Ly-$\alpha$ emission line, however, both the emission line
and absorption profiles are strikingly different.  The sightline to \UMB\ has
a larger absorbing column density and/or a higher covering fraction in H\,I
and N\,V.  The Ly-$\alpha$ emission line in \UMB\ is at least a factor of two
greater in equivalent width than in \UMA.  

 Large spectral differences have been noted in accepted lensed
systems, so do not rule out the lens hypothesis.  SBS~1520+530  is a
lensed BALQSO with a detected lens galaxy and time delay, and  
the two components show significantly different emission line
equivalent widths \citep{burud02b}.  This could be due to differential
amplification of the continuum emitting region caused by microlensing.
HE~2149-2745 \citep{burud02a} is another lensed BALQSO system with a detected
time-delay; it also shows large equivalent width differences in the
emission lines, but none in the BALs.  The continuum slopes are
different, due either to microlensing or reddening by the (as-yet
unidentified) lens galaxy.  

Within the lens hypothesis, spectral differences could also be caused by
pathlength time-delays combined with spectral variability.  In this case we are
effectively viewing one quasar at two epochs, so this difference could
plausibly be explained by a combination of intrinsic emission line variability
\citep[see e.g.][]{obrien91,small97} coupled with different, possibly variable,
absorption profiles \citep{mich96}.

Under the binary hypothesis, the different Ly-$\alpha$ equivalent
widths should be consistent with the global Baldwin effect \citep[an
anti-correlation of line 
equivalent width with luminosity observed in quasar samples;][]{baldwin77}.
From the slope of $\beta=-0.14 \pm 0.02,$ (where $W_{\lambda}$(Ly$\alpha$)
$\propto L_{\lambda 1450}^{\beta}$) derived by \citet{dietrich02}, the factor
of $\sim$100 difference in brightness corresponds to an expected equivalent
width that is about twice as large in \UMB.  Within the uncertainties
caused by the strong absorption bands and the large scatter in the
Baldwin relationship, this ratio is consistent with the observed STIS spectra.


\section{HST imaging}
\begin{figure*}
\centerline{
\resizebox{3.2in}{!}{\includegraphics{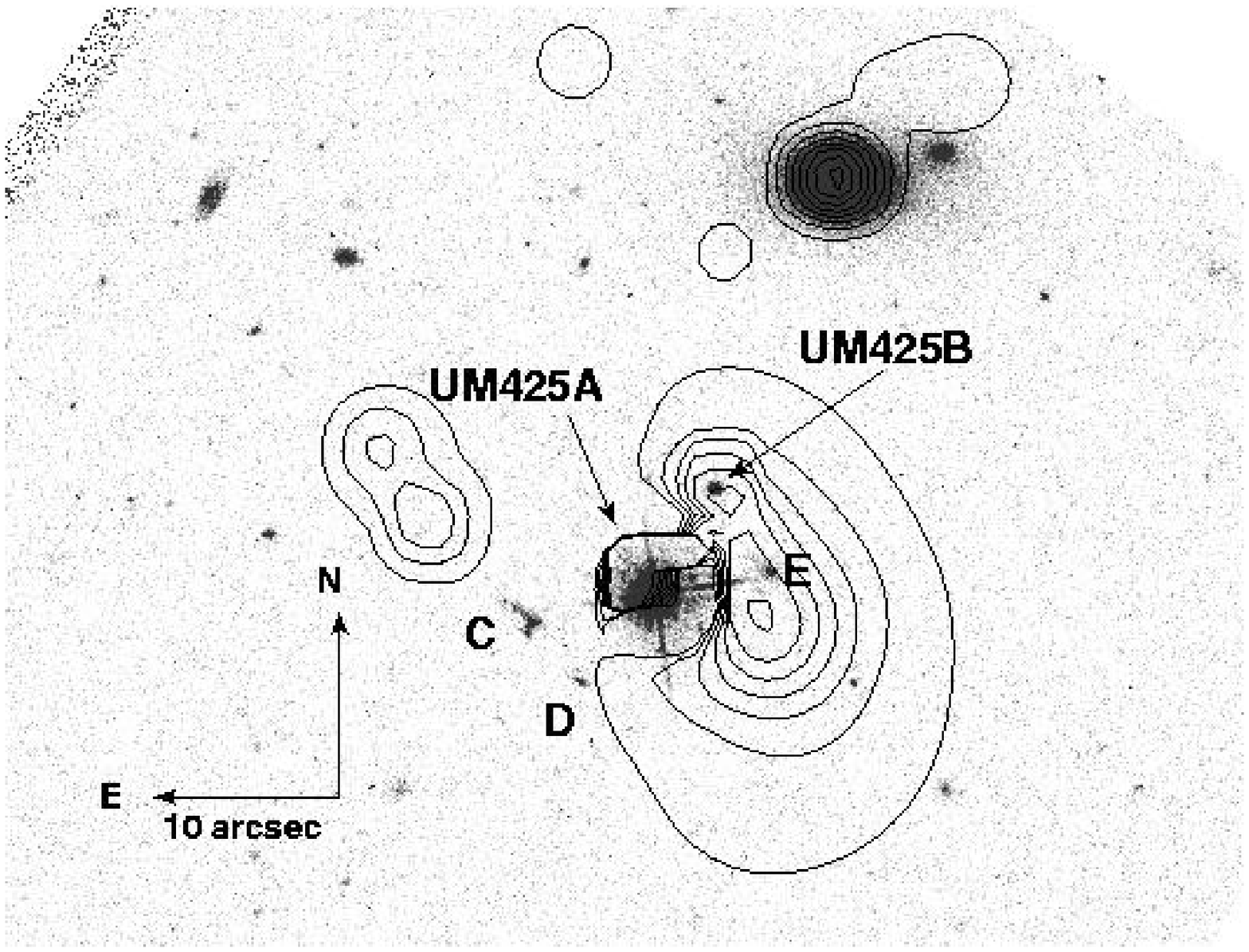}}
\resizebox{3.2in}{!}{\includegraphics{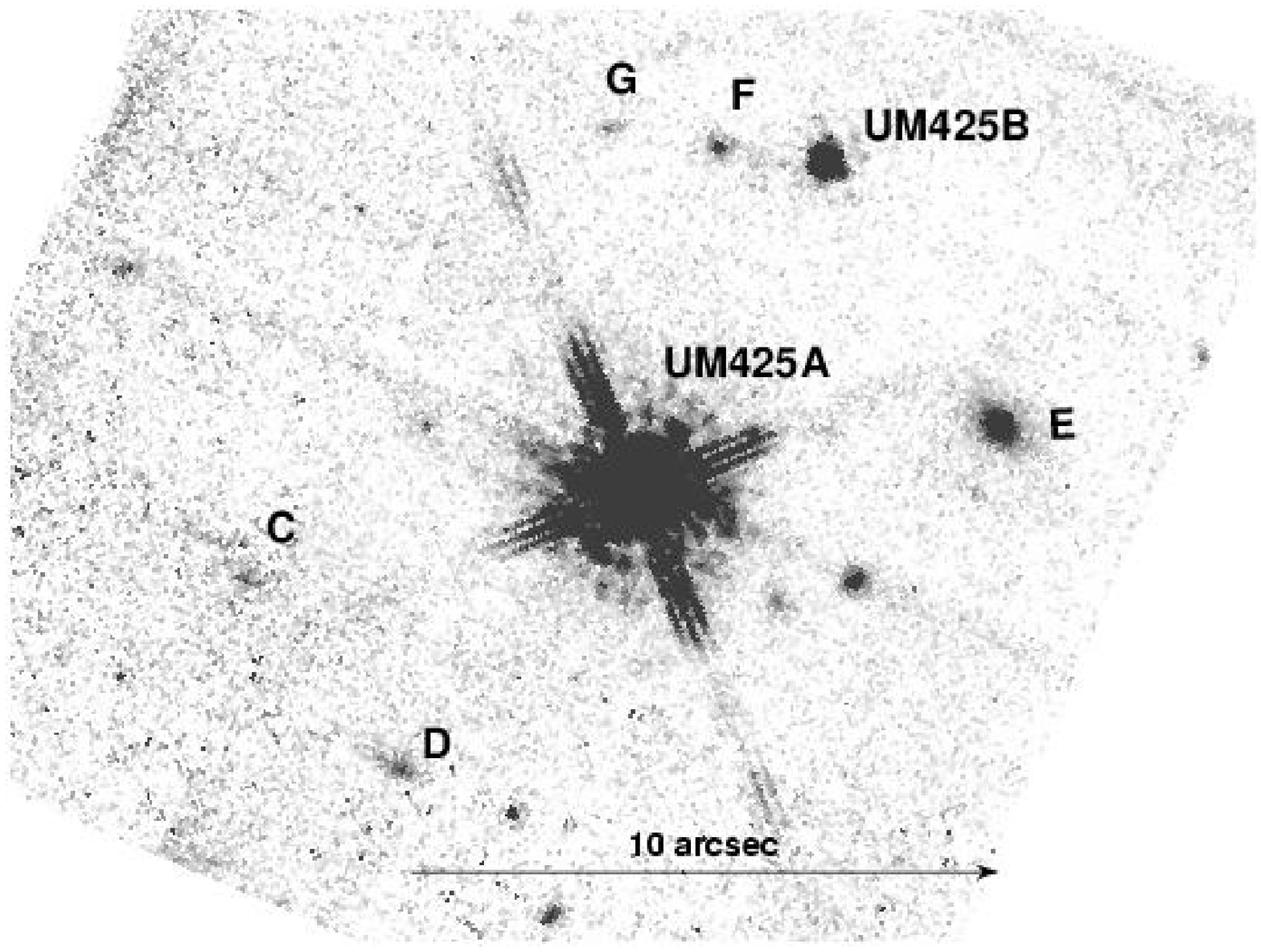}}
}
\caption{Left panel: HST/WFPC image of the \UM\ field overlayed with
contours from our PSF-subtracted \Chandra/ACIS image
(Figure~\ref{fig:diffuse_emission}).  Right panel: HST/NICMOS image of
the \UM\ field.  In both panels the object labels A to E correspond to
those defined in \citet{meylan89} and \citet{courbin95}. }

\label{fig:wfpc_nicmos}
\end{figure*}

The \UM\ field was imaged with HST/WFPC2 (PI J. Westphal) on 1995-May-01 for
600~sec with the F555W filter and for 1400~sec with the F814W filter.  On
1998-May-28, the field was imaged in the infrared for 2560~sec using HST/NICMOS
\citep[imaged as part of the CASTLES project; ][]{munoz98} with the F160W
filter.  We retrieved the data from the HST
archive and used the standard IRAF tasks \texttt{mscimage} and
\texttt{imcombine} to align and combine the individual exposures into
a single final image.  For the NICMOS data we removed a pedestal bias
variation using the STSDAS tool \texttt{pedsky}.  The reduced WFPC2
and NICMOS images are shown in Figure~\ref{fig:wfpc_nicmos}.  The left
panel shows the WFPC2 image overlayed with contours from our
PSF-subtracted ACIS image (Figure~\ref{fig:diffuse_emission}), while
in the right panel we show the NICMOS image.  

In both panels the
object labels A to E correspond to those defined in \citet{meylan89}
and \citet{courbin95}.  The faint objects F and G are also seen in the
the WFPC2 image, but were not previously identified in ground-based
imaging.  The contours show the X-ray emission, including the diffuse
component centered about 5\arcsec\ west of \UMA, the point-like
emission from \UMB, emission to the north from a bright foreground
galaxy ($V \approx 17.8$) at $z=0.1265$ \citep{meylan89}, and possible
optical blank-field sources about 10\arcsec\ to the east.

The NICMOS and WFPC images immediately confirm that the objects C, D,
and E are extended.  Object C has a notably disturbed morphology and
is likely in the process of a merger. \citet{meylan89} noted the large
number of faint galaxies in the field, which taken together with the
slight spectroscopic difference of \UMA\ and \UMB, led them to
postulate the existence of cluster at $z \approx 0.6$.  Within a
radius of 10\arcsec\ of \UMA\ we see no fewer than 10 confirmed
sources which appear in both the NICMOS and WFPC images.  This stands
in stark contrast to another well-studied wide separation quasar pair
Q1634+267A,B in which NICMOS imaging (2560~sec with the F160W filter)
revealed no non-stellar sources within 8\arcsec\ of the pair 
\citep{peng99}.

To estimate the significance of the apparent overdensity of
sources in the NICMOS image, we have calculated $H$ band magnitudes
for all sources within a 16\arcsec$\times$16\arcsec\ box centered on
\UMA.  This was done using aperture photometry to calculate the net
counts (DN~s$^{-1}$) for all sources which were detected in both the
WFPC and NICMOS images.
We converted the net count rate to a flux at 1.6$\mu$m using the
inverse sensitivity factor
$2.406\times10^{-19}$~ergs\,cm$^{-2}$\,\AA$^{-1}$\,DN$^{-1}$ supplied in the
calibrated NICMOS image FITS file header.  Finally, the flux was
converted to an $H$ magnitude using the NICMOS Units Conversion
Form\footnote{http://www.stsci.edu/hst/nicmos/tools/conversion\_form.html}.
For the two quasars \UMA\ and \UMB\ we find $H$ magnitudes of 14.1 and
18.6~mag, respectively.  The magnitude difference of 4.5~mag is
consistent with the average R-band difference of 4.4~mag seen by
\citet{courbin95}.

A total of 11 objects were processed in this way, with $H$ magnitudes
as faint as 22.7~mag.  Excluding \UMA\ and \UMB\ (since they are known
background objects and the field was selected for their presence),
there are 6 objects brighter than $H = 22.0$.  At levels fainter than
this our source detection becomes incomplete.  We calculate the
expected number of sources by using the NICMOS H-band number count
versus magnitude given by \citet{yan98}.  Integrating the number count
relation up to a faint limit of 22.0~mag gives a value of 20.3 sources
per square arcmin.  In our 16\arcsec$\times$16\arcsec\ box, with a
2\arcsec\ radius around \UMA\ excluded, we therefore expect 1.38
sources.  The probability of seeing 6 or more sources is 0.0030,
implying that there is an overdensity of galaxies in the \UM\ field at
approximately 3-$\sigma$ confidence.  For \UM\ the combination of
diffuse X-ray emission and a rich field of galaxies strongly suggests
the presence of a galaxy group or cluster in this direction.
We discuss this possibility further in \S~\ref{sec:discussion}.

If \UM\, were lensed, a lensing galaxy would be expected nearer the faint
component.  We see no evidence for a possible lensing galaxy near
\UMB.  Taking the image of \UMA\ to define the instrument PSF, we used
the CIAO Sherpa fitting program to subtract the PSF from the image of \UMB.
Outside a core radius of 0.15\arcsec\ we saw no significant residuals.
We then created simulated images by adding the source counts from one
of the faint ($H=21.7$~mag) galaxies in the field to the \UMB\ image.
This was done for several positions on the line between \UMB\ and
\UMA, and we found we would clearly detect such a galaxy at a distance
greater than 0.3\arcsec\ from \UMB.  A bright galaxy such as the
$H=19.9$ galaxy in the field would be detected even if it were exactly
coincident with \UMB.  



\section{Discussion}
\label{sec:discussion}

\subsection{A Galaxy Group or Cluster as a Lens Candidate}

If a wide-separation quasar pair (WSQP) is produced by a lens mass modeled as a
simple, singular
isothermal sphere \citep[SIS, see][]{schneider92}, the most 
conservative flux limits are derived by assuming the lens lies at the
``minimum flux redshift,'' the redshift that would minimize the
observed X-ray flux.  If we neglect K-corrections, the flux from the
lens is \begin{equation}
  F =  { L \over 4\pi D_{OL}^2(1+z_l)^2 }  \propto
    \left[ { D_{OS} r_H \over D_{OL} D_{LS} (1+z_l) } \right]^2
\end{equation}
\noindent
where $r_H$ is the Hubble radius $c/H_0$.  For our assumed cosmology,
the flux is minimized at a lens redshift of $z_l=0.6$. 

The image separation $\Delta\theta= 8\pi(\sigma_v/c)^2 D_{LS}/D_{OS}$
depends only on the velocity dispersion of the potential $\sigma_v$ 
and the ratio of the comoving distances\footnote{We calculate angular
size distances in our cosmological model using the {\tt ANGSIZ} code
of \citet{kayser97}.} between the lens and the 
source, $D_{LS}$, and the observer and the source, $D_{OS}$. 
In the SIS model for the lensing mass, and using the minimum flux
redshift for $z_{l}$, the observed image separation of
6$\arcsec$.5 implies a cluster velocity dispersion of $\sigma_v=480
\kms$.  This corresponds to a {\em minimum} enclosed cluster mass of
$5.4\times 10^{13}\msun$ to induce the observed pair separation.
Combining the $L_X-\sigma_v$ relation from, 
e.g., \citet{mulchaey98} and the $L_X-T$ relation from,
e.g., \citet{markevitch98}, and neglecting any possible cosmological
evolution of these relations for a qualitative estimate, we obtain
$L_X (0.1-2.4~{\rm keV})\approx 1.5\times 10^{43}$~\lcgs\, and $kT\approx
1.5$~keV for such a cluster.  At $z=0.6$, this corresponds to $f_X
(0.1-2.4~{\rm keV})\approx 10^{-14}$~\fcgs.  The flux we observe 
is about 6 times fainter than this, so {\em we detect no normal cluster 
or group that could be solely responsible for the observed image
splitting in a lens scenario.}

Do the galaxies detected in the WFPC2 and NICMOS images suggest
the existence of a group at the minimum flux redshift?  We use the
public {\em HyperZ} photometric redshift code of \citet{hyperz00},
where an E/S0 spectral energy distribution yields $H-K=0.7$.
Together with a characteristic magnitude of $M_K^*= -24.75$
\citep{gardner97, glazebrook95} we take $M_H^*\sim= -24.05$.  
The brightest galaxy in the near field (marked E in
Figure~\ref{fig:wfpc_nicmos}) has $H=19.9$, which if at $z\sim0.6$
corresponds to 0.3$L_H^*$, a fairly bright galaxy.  Most of the
objects detected near the NICMOS flux limit correspond to about
0.04$L_H^*$, and would be small galaxies at this
redshift.  Therefore, these objects are plausible members of a group
or small cluster at $z\sim 0.6$.  If these objects are galaxies in a
lensing group at a higher redshift, then its X-ray flux is more than a
factor $\sim 5$ lower for its mass than seen in nearby groups, which
could imply a baryon fraction at most half normal.  

In lenses clearly due to a combination of a cluster and a galaxy
\citep[particularly Q0957+561;][]{keeton00}, a massive, luminous lens
galaxy dominates the image splitting.  Here we see no such candidate
galaxy, even in the infrared, to a limit of approximately
$L_*/20$.  Such a galaxy, unless completely different from all other
known lens galaxies \citep[e.g.,][]{rusin03,kochanek00,xanthopoulos98},  
must make a negligible contribution to the overall image separation.

\subsection{Optical Brightness Argues for a Lens}

\begin{figure*}
\centerline{
 \resizebox{3in}{!}{\includegraphics{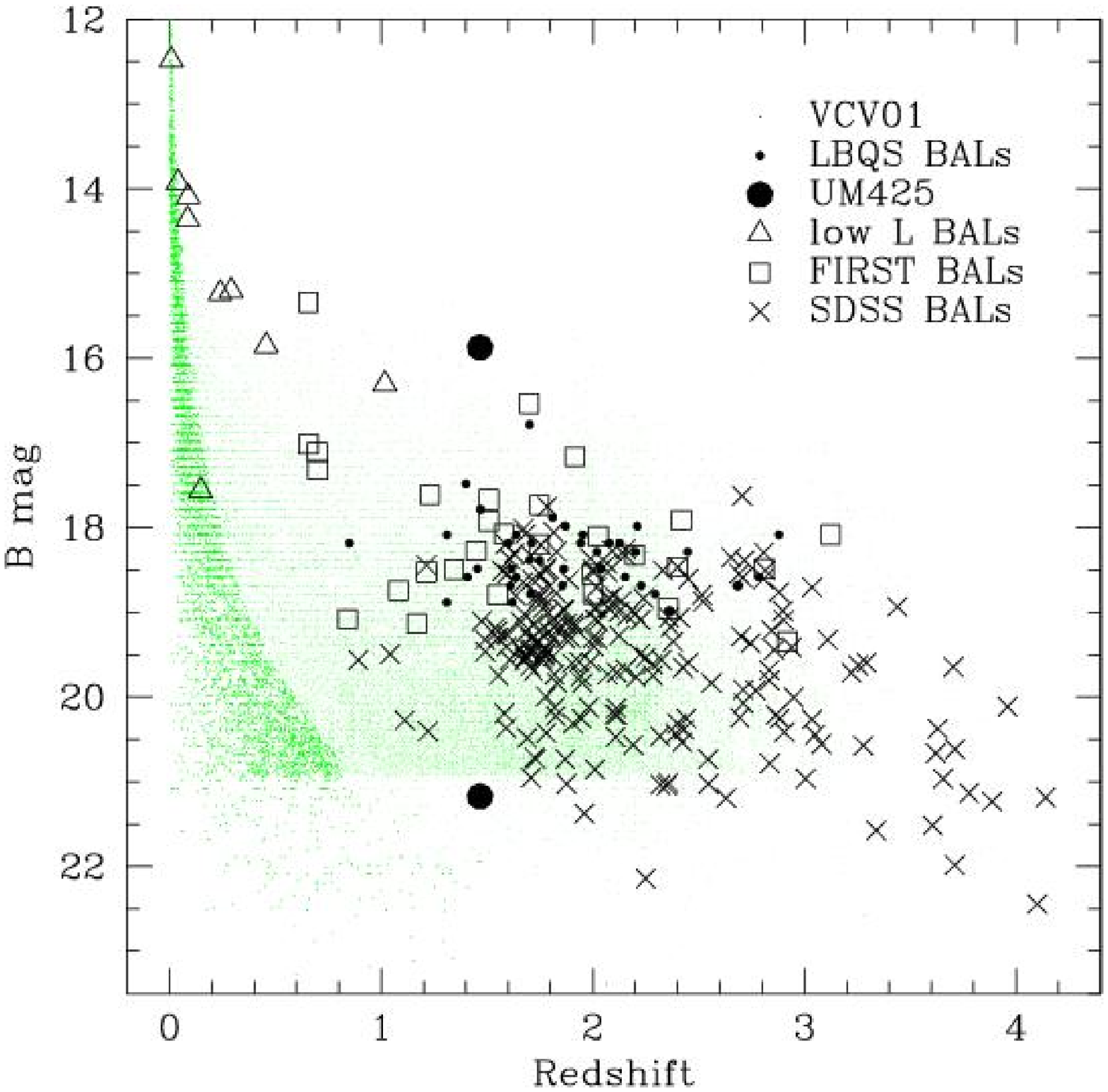}}
 \resizebox{3in}{!}{\includegraphics{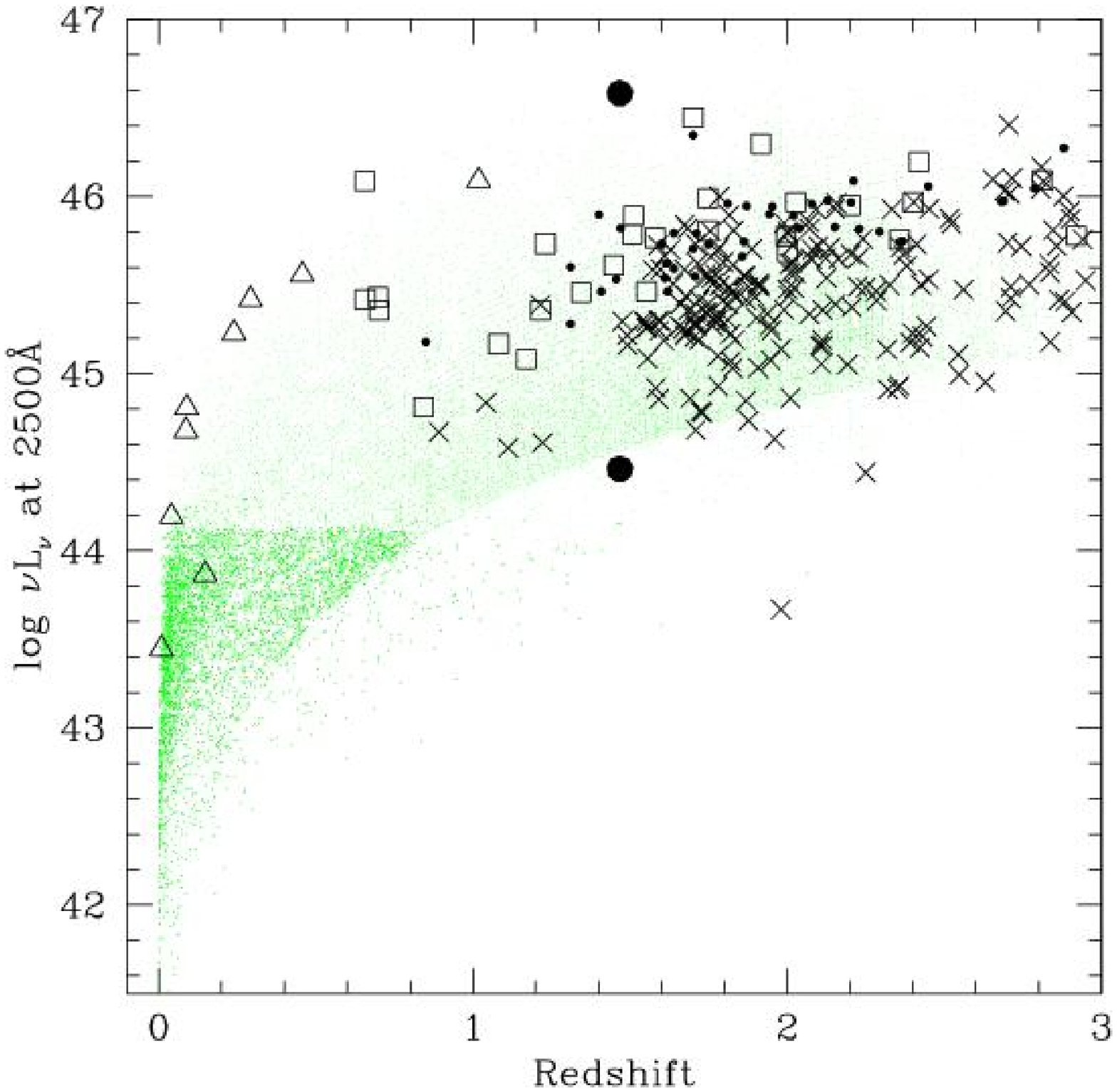}}
}
\caption{Optical (blue) magnitude (Left panel) and luminosity at
2500\,\AA\ (Right panel) versus redshift.  Small dots show these quantities
for about 27,000 quasars and active galaxies from 
\citet{veron01}. BALQSOs in the SDSS, LBQS, and FIRST surveys
are marked with large open symbols, as are 9 low redshift AGN with
BALs or mini-BALs taken from the literature. The large filled
circles are data points for \UMA\ and \UMB.  \UMA\ is anomalously
bright compared to most quasars, which is how it was originally
selected as a lens candidate \citep{meylan89}.  \UMB\ is anomalously
dim for a BALQSO. } 
\label{fig:opt_lum}
\end{figure*}

\UMA\ is about an order of magnitude brighter (in the optical) than
quasars at comparable redshifts.  Figure~\ref{fig:opt_lum}
compares its luminosity to 27,000 quasars and AGN from
\citet{veron01}.  This observation would suggest that \UMA\ might be
magnified by a lens; anomalous brightness was exactly the criterion
that \citep{meylan89} used originally to select the object as a lens
candidate \citep{meylan89}.  \UMA\ is especially bright for a
BALQSO. Based on polarization studies \citep{goodrich97}, BALQSO
fluxes may be attenuated by a factor of about 5, which contributes to
their diminished representation in flux-limited optical surveys
\citep{hewett03}.  Figure~\ref{fig:opt_lum} shows BALQSOs from   
the Large Bright Quasar \citep{hewett95}, the SDSS \citep{reichard03}, 
and FIRST Bright Quasar \citep{becker00} surveys.

Figure~\ref{fig:opt_lum} also shows that \UMB\ is underluminous for a
BALQSO.  This further supports a lensing interpretation because to
date, because strong BALs like these are found only in luminous QSOs.  This is
partly a selection effect, because CIV BALs do not enter the
observed-frame optical bandpass until $z\sim 1.3$ and 
a CIV BALQSO as faint as \UMB\ would only have been detected
in a ground-based optical survey if it extended to $B\sim 22$.  
To find CIV BALs in nearby low-luminosity AGN, a large dedicated UV
spectroscopic  survey is required (e.g., GALEX; \citealt{martin97}).
However, among the hundreds of existing UV spectra of lower luminosity
($M_B>-23$) AGN, not a single example of a strong BAL is known.

The lowest luminosity BAL analogs have been found from space-based UV
spectra of known AGN, or from the  less common low-ionization Mg\,II
BALs which enter the optical band for $z\gapprox 0.3$.  A thorough
search of the literature reveals just a handful of $z\lapprox 1$
objects with broad absorption lines.  The low redshift BAL AGN are
shown as open triangles in Figure~\ref{fig:opt_lum}. We could find
only 3 BALQSOs less luminous than \UMB, all at $z<0.2$.  All these
indeed fall into two classes. 
(1) The loBAL QSOs.  These loBALs constitute a much
higher fraction (27\%) of IR-selected than optically-selected QSOs
(1.4\%; \citealt{boroson92}), and are thought to provide evidence for
a link between interaction/merger processes and accretion
\citep{canalizo01}.  The FeloBALs like IRAS07598+6508
\citep{hines95} or SDSS\,173049.10$+$585059.5 \citep{hall02}
are particularly highly reddened and
polarized and show absorption from excited fine-structure levels or
excited atomic terms of Fe\,II or Fe\,III \citep{becker97}.  However,
\UMB\ shows neither spectroscopic (see Figure~\ref{fig:stis}) nor
photometric evidence for being a loBAL since the colors of \UMA\ and
\UMB\ are identical within the errors \citep{meylan89}.
(2) The so-called 'mini-BALs' such as PG~1115+080,  PG~1411+442,
RX~J0911.4+0551, or NGC3516 (e.g., \citealt{kraemer02}).  These may be
the low luminosity analogs of BALQSOs, but have weaker BALs.  From the
accepted CIV `balnicity' test of \citet{weymann91} might also
classify \UM, as a miniBAL, but the low S/N blueward of CIV 
in the spectrum of \citet{michalitsianos97}, makes it
difficult to tell.  The balnicity test is not well-defined for
lines other than CIV, but the Ly$\alpha$ and NV absorbers in
Figure~\ref{fig:stis} are impressive enough (we measure equivalent
widths of $32\pm5$ and $42\pm5$\AA, respectively).

True BALs are virtually unknown in low-luminosity quasars.
Correlations in absorbed (soft X-ray weak) quasars between
luminosity  and CIV absorption (both equivalent width and maximum
outflow velocity) suggest that this trend may hold up in future
samples as well \citep{laor02}. 

\UMA\ is quite bright for a QSO, and especially for a
BALQSO, so may be magnified by a lens.  \UMB\ is quite
dim for a BALQSO, so may be demagnified by a lens.  In summary, the
anomalous brightnesses of both \UMA\ and \UMB\ argue for the lens
interpretation.  

\subsection{Implications of Lensing for BAL structure}

Standard lensing models may not be easily applicable in the case
of BALQSOs, if there is structure to the outflowing clouds on angular
scales similar to the image separation.  For instance, direct
application of the observed optical flux ratios may be misleading,
since our \Chandra\  observation already demonstrates the likelihood of
significantly different absorbing columns along the line of sight,
and the relative fractions of transmitted, obscured, and reflected 
light are unknown.

If \UM\ is lensed, then the two components image slightly different
sightlines from the lens to the QSO central continuum source, probing
potentially different absorbing material.  If the lens is roughly half the
proper distance to the quasar, the sightlines will be be separated by
$\sim 6$\arcsec.5.  At a radius of 5~pc, the sightlines would have a
transverse separation of $5\times 10^{14}$~cm.
Probing the structure of the BAL medium at this size scale is quite
interesting in the context of quasar structure, and particularly for
recent multiphase models for the BAL outflow, such as the model
proposed by \citet{everett02}.  These authors are able 
to fit high-resolution optical spectroscopic data for the BALQSO
FIRST~J1044+3656 with a dense cloud ($n = 10^{8.5}$~cm$^{-3}$) at a
distance of $r \sim 4$~pc, embedded in a warm outflowing medium.  If a
similar physical situation applies for \UM, and the differential X-ray
absorbing column $N_{H,\rm UM425B} - N_{H,\rm UM425A}
=1.5\times 10^{23}$~cm$^{-2}$ were due to a single cloud, then that
cloud would have a linear dimension a roughly $5\times 10^{14}$~cm.
This nicely matches the transverse separation of the two sightlines.

A complication to this reasoning is that the gravitational time delay of at
least 1.7~years \citep[e.g.][]{michalitsianos97} means that we are viewing the
two sightlines at different epochs.  In that time (observed frame) an absorbing
cloud would travel $2.1\times 10^{16}$~cm along the line of sight, assuming an
outflowing velocity of 10000~km\,sec$^{-1}$.  Unless the absorbing cloud
velocities are directed very nearly toward our line of sight, the geometric
distance between the two lines of sight is overwhelmed by the effective
temporal distance due to the outflow velocity and the lens time delay.
Nevertheless, it should be noted that BAL models which assume a wind that is
radiatively-driven from the central source will naturally produce velocities
that are primarily radial.  The magnitude of the residual transverse component
(e.g. if the wind is launched from the disk) is model dependent.  It could be
the case that the geometric and temporal distances are of comparable size.


\subsection{Interaction/Merger Hypothesis}

The {\em a priori} odds of finding a BALQSO pair such as this seems
low even using more recent estimates of the true fraction of
BALQSOs of 20-33\% \citep{hewett03, reichard03}.  However, the close
interaction or ongoing merger of such a close pair may enhance
the probability if outflow (BAL) activity is triggered in both
nuclei.  If not lensed, \UM\ would represent evidence 
for merger-induced BALQSO activity, giving strength to arguments that
the BAL phenomenon may be an evolutionary phase of QSO activity.
A large sample of WSQPs could serve to constrain the lifetimes
of triggered activity.  As an example, the projected linear separation
of the pair is $\sim 40$~kpc.  Assuming they have relative velocities 
of about 600\kms\, \citep{michalitsianos97}, the induced activity has
already lasted of order 70~Myr.

\section{Summary}

In summary, the arguments for a lens interpretation are:
\begin{itemize}
	\item similar redshift, colors, and spectral type of \UMA\ and \UMB
	\item anomalous brightness of \UMA\, for a quasar at $z=1.465$
	\item anomalous faintness of \UMB\, for a BALQSO
	\item some X-ray and $H$-band evidence for an
	intervening galaxy group 
\end{itemize}

The arguments for the binary hypothesis are:
\begin{itemize}
	\item significantly different X-ray absorption
	\item any putative intervening galaxy group 
is $6\times$ X-ray underluminous for the required lensing mass
	\item significantly different UV emission line profiles
	\item emission line equivalent widths consistent with the
Baldwin Effect for the apparent luminosity differences
\end{itemize}

Unfortunately, while the evidence weighs toward the binary hypothesis,
resolution of the debate is not final.  The UV spectroscopic 
differences are similar to those seen in some definitively lensed quasars,
and larger absorption column differences in X-rays than UV are
quite plausible given that the size, placement and ionization state of
quasar X-ray and UV absorbing regions probably differ (e.g.,
\citealt{krolik01, hamann01}).  As often noted, the  most definitive
test is detection of a time delayed variability between the 
image components.  For a given image separation in the lens scenario,
the time delay depends strongly on the asymmetry $(r_A-r_B)/(r_A+r_B)$
of the images with respect to the lens, with the time delay smallest
for a symmetric system.  Here the lens position is unknown.  For large 
flux ratios, the asymmetry is assumed to be large so that the
asymmetry factor goes to unity, and the time delay is maximized
(its value depending mostly on the image separation $r_{AB}$).  
If the maximal delay of $\sim 10$ years holds \citep{courbin95},
photometric variability studies are prohibitively difficult.
A more efficient method for proving the existence of lensing is near infrared
imaging detection of the quasar host galaxy morphology with sufficient
resolution and signal to see shear (e.g., FSC10214+4724; \citealt{evans99},
B0712+472; \citealt{jackson00}).

We wish to thank Aneta Siemiginowska for helpful discussions on
lensing and the 
extended emission, and Smita Mathur for comments on and contributions to the
original \Chandra\  proposal.  Thanks to Chris Kochanek and Josh Winn for
guidance and comments.  This work was supported by CXO grant GO2-3132X
and NASA grant NAS8-39073. PJG and TLA gratefully acknowledge support
through NASA  Contract NASA contract NAS8-39073 (CXC).

{}

\clearpage

\clearpage
\clearpage

\clearpage

\clearpage

\clearpage

\clearpage

\clearpage

\clearpage

\clearpage


\begin{thebibliography}{}


\bibitem[Arav et al.(2001)]{arav01} Arav, N. et al. 2001, ApJ, 561, 118

\bibitem[Avni \& Tananbaum(1982)]{avni82} Avni, Y., \& Tananbaum, H. 1982, ApJ, 262, L17

\bibitem[Baldwin(1977)]{baldwin77} Baldwin, J.A. 1977, ApJ, 214, 679

\bibitem[Becker et al.(1997)]{becker97} Becker, R.~H., Gregg, 
M.~D., Hook, I.~M., McMahon, R.~G., White, R.~L., \& Helfand, D.~J.\ 1997, 
\apjl, 479, L93 

\bibitem[Becker et al.(2000)]{becker00} Becker, R.~H.,  White, R.~L.,
Gregg, M.~D., Brotherton, M. S., Laurent-Muehleisen, S.~a.,
\& Arav, N. 2000, \apj, 538, 72

\bibitem[Bolzonella, Miralles, \& Pello(2000)]{hyperz00}
Bolzonella, M., Miralles, J.-M., \& Pello, R. 2000, A\&A 363, 476

\bibitem[Boroson \& Meyers(1992)]{boroson92} Boroson, T.~A.~\& 
Meyers, K.~A.\ 1992, \apj, 397, 442 

\bibitem[Brandt \& Gallagher(2000)]{brandt00} Brandt, W. N. \& Gallagher,
S. C. 2000, NewAR, 44, 461   

\bibitem[Burud et al.(2002a)]{burud02a} Burud, I. et al. 2002, A\&A,
  383, 71

\bibitem[Burud et al.(2002b)]{burud02b} Burud, I. et al. 2002, A\&A,
  391, 481


\bibitem[Canalizo \& Stockton(2001)]{canalizo01} Canalizo, G.~\& 
Stockton, A.\ 2001, \apj, 555, 719 

\bibitem[Chartas et al.(2002)]{chartas02} Chartas, G.~et al. 2002, \apj, 579. 169

\bibitem[Courbin et al.(1995)]{courbin95} Courbin, F.~et al.\ 
1995, \aap, 303, 1 

\bibitem[Davis(2001)]{davis01} Davis, J.~E.\ 2001, \apj, 562, 
575 

\bibitem[Dietrich et al.(2002)]{dietrich02} Dietrich, M., Hamann, 
F., Shields, J.~C., Constantin, A., Vestergaard, M., Chaffee, F., Foltz, 
C.~B., \& Junkkarinen, V.~T.\ 2002, \apj, 581, 912 

\bibitem[Egami et al.(1996)]{egami96} Egami, E., Iwamuro, F., Maihara,
T., Oya, S., \& Cowie, L. L., 1996, AJ 112, 73

\bibitem[Elvis(2000)]{elvis00} Elvis, M.\ 2000, \apj, 545, 63 

\bibitem[Evans et al.(1999)]{evans99} Evans, A.~S., Scoville, 
N.~Z., Dinshaw, N., Armus, L., Soifer, B.~T., Neugebauer, G., \& Rieke, M.\ 
1999, \apj, 518, 145 

\bibitem[Everett, K{\" o}nigl, \& Arav(2002)]{everett02} Everett,
J., K{\" o}nigl, A., \& Arav, N.\ 2002, \apj, 569, 671

\bibitem[Fabian(1999)]{fabian99} Fabian A. 1999, MNRAS, 308, L39

\bibitem[Fabian et al.(2001)]{fabian01} Fabian A.~C,
Crawford, C.~S., Ettori, S., \& Sanders, J.~S. 2001, MNRAS, 322, L11

\bibitem[Fan et al.(2003)]{fan03} Fan, X. et al. AJ, 2003, in press

\bibitem[Gallagher et al.(2002a)]{gallagher02a}Gallagher, S. C., Brandt,
W. N., Chartas, G., Garmire, G. P 2002, ApJ, 567, 37

\bibitem[Gallagher et al.(2002b)]{gallagher02b} Gallagher, S.~C., 
Brandt, W.~N., Chartas, G., Garmire, G.~P., \& Sambruna, R.~M.\ 2002, \apj, 
569, 655 

\bibitem[Gardner et al.(1997)]{gardner97}Gardner, J. P., Sharples, R. M., Frenk, C. S., 
Baugh, C. M., \& Carrasco, B. E. 1997, \apj, 480, L99 

\bibitem[Glazebrook et al.(1995)]{glazebrook95}Glazebrook, K.,
Peacock, J. A., Miller, L., \& Collins, C. A. 1995, \mnras, 275, 169  

\bibitem[Goodrich(1997)]{goodrich97} Goodrich, R.~W. 1997, ApJ, 474, 606

\bibitem[Green et al.(1995)]{green95} Green, P.~J.~et al.\ 
1995, \apj, 450, 51 

\bibitem[Green et al.(2001)]{green01} Green, P. J. et al 2001, ApJ,
558, 109

\bibitem[Green et al.(2003)]{green03} Green, P. J. et al 2003, ApJ, submitted

\bibitem[Gregg et al.(2002)]{gregg02} Gregg, M.~D., Becker, R.~H.,
  White, R.~L.,  Richards, G.~T., Chaffee, F.~H., \& Fan,~X.\ 2002,  
\apjl, 573, 85


\bibitem[Hall et al.(2002)]{hall02} Hall, P.~B.~et al.\ 2002, 
\apjs, 141, 267 

\bibitem[Hamann \& Ferland(1993)]{hamann93} Hamann, F.~W., \& Ferland,
G.J. 1993, ApJ, 418, 11

\bibitem[Hamann \& Ferland(1999)]{hamann99} Hamann, F.~W., \& Ferland, G. 
1999, ARA\&A, 37, 487


\bibitem[Hamann, Netzer, \& Shields(2001)]{hamann01} Hamann, F.~W.
Netzer, H., \& Shields, J.~C. 2001, ApJ, 436, 101

\bibitem[Hasinger, Schartel, \& Komossa(2002)]{hasinger02} 
Hasinger, G., Schartel, N., \& Komossa, S.\ 2002, \apjl, 573, L77 

\bibitem[Hazard et al.(1984)]{hazard84} Hazard, C., Morton, D. C.,
Terlevich, R., \& McMahon, R. 1984, ApJ, 282, 33

\bibitem[Hewett, Foltz,  \& Chaffee(1995)]{hewett95}
Hewett, P.~C., Foltz, C.~B., \& Chaffee, F.~H. 1995, AJ, 109, 1498

\bibitem[Hewett \& Foltz(2003)]{hewett03} Hewett, P. C. \& Foltz,
C. B. AJ, 2003, in press (astro-ph/0301191)

\bibitem[Hines \& Wills(1995)]{hines95} Hines, D.~C.~\& Wills, 
B.~J.\ 1995, \apjl, 448, L69 

\bibitem[Jackson, Xanthopoulos, \& Browne(2000)]{jackson00} 
Jackson, N., Xanthopoulos, E., \& Browne, I.~W.~A.\ 2000, \mnras, 311, 389 

\bibitem[Kayser, Helbig, \& Schramm(1997)]{kayser97}
Kayser, R., Helbig, P., \& Schramm, T., 1997, A\&A, 318, 680

\bibitem[Keeton et al.(2000)]{keeton00} Keeton, C.~R.~et al.\ 
2000, \apj, 542, 74 

\bibitem[Kochanek et al.(2000)]{kochanek00} Kochanek, C.~S.~et 
al.\ 2000, \apj, 543, 131 

\bibitem[Korista et al.(1996)]{korista96} Korista, K. T. et al. 1996, ApJ, 461, 641

\bibitem[Kraemer et al.(2002)]{kraemer02} Kraemer, S.~B., 
Crenshaw, D.~M., George, I.~M., Netzer, H., Turner, T.~J., \& Gabel, J.~R.\ 
2002, \apj, 577, 98 


\bibitem[Krolik \& Kriss(2001)]{krolik01} Krolik, J.~H., \& Kriss,
  G.~A. 2001, ApJ, 561, 684


\bibitem[Laor \& Brandt(2002)]{laor02} Laor, A., \& Brandt, W. N. 2002, \apjl, 569, L641

\bibitem[Lewis et al.(2002)]{lewis02} Lewis, G. F.,
Ibata, R. A.; Ellison, S. L., Aracil, B., Petitjean, P.,
Pettini, M., \& Srianand, R. 2002, MNRAS, 334, L7

\bibitem[Markevitch(1998)]{markevitch98} Markevitch, M.\ 1998, \apj, 
504, 27 

\bibitem[Martin et al.(1997)]{martin97} Martin, C.~et al.\ 1997, 
Bulletin of the American Astronomical Society, 29, 1309 

\bibitem[Mathur(2000)]{mathur00} Mathur, S.\ 2000, \mnras, 314, 
L17 

\bibitem[Mathur \& Williams(2003)]{mathur03} Mathur, S., \& Williams,
 R. J. \ 2003, ApJL, submitted 

\bibitem[Meylan \& Djorgovski(1989)]{meylan89} Meylan, G.~\& 
Djorgovski, S.\ 1989, \apjl, 338, L1 

\bibitem[Michalitsianos, Oliversen, \&  Maran(1996)]{mich96}
 Michalitsianos, A.~G.,  Oliversen, R. J. \&  Maran, S. P.\ 1996,
 \apjl, 458, 67  

\bibitem[Michalitsianos, Falco, Munoz, \& 
Kazanas(1997)]{michalitsianos97} Michalitsianos, A.~G., Falco, E.~E., 
Mu{\~ n}oz, J.~A., \& Kazanas, D.\ 1997, \apjl, 487, L117 

\bibitem[Mulchaey \& Zabludoff(1998)]{mulchaey98} Mulchaey, 
J.~S.~\& Zabludoff, A.~I.\ 1998, \apj, 496, 73 

\bibitem[Mu{\~ n}oz et al.(1998)]{munoz98} Mu{\~ n}oz, J.~A., 
Falco, E.~E., Kochanek, C.~S., Leh{\' a}r, J., McLeod, B.~A., Impey, C.~D., 
Rix, H.-W., \& Peng, C.~Y.\ 1998, \apss, 263, 51 

\bibitem[O'Brien \& Gondhalekar(1991)]{obrien91} O'Brien, 
P.~T.~\& Gondhalekar, P.~M.\ 1991, \mnras, 250, 377 

\bibitem[Peng et al.(1999)]{peng99} Peng, C.~Y.~et al.\ 1999, 
\apj, 524, 572 

\bibitem[Reeves \& Turner(2000)]{reeves00} Reeves, J.~N.~\& 
Turner, M.~J.~L.\ 2000, \mnras, 316, 234 

\bibitem[Reichard et al.(2003)]{reichard03} Reichard, T. A. et al.
2003, \aj, in press (astro-ph/0301019)

\bibitem[Rusin et al.(2003)]{rusin03} Rusin, D. et al. 2003, \apj, in press
(astro-ph/0211229)

\bibitem[Scargle(2003)]{scargle03} Scargle, J.~D.\ 2003, {\em in prep}

\bibitem[Scargle(1998)]{scargle98} Scargle, J.~D.\ 1998, \apj, 
504, 405 

\bibitem[Schmidt \& Hines(1999)]{schmidt99} Schmidt, G. \& Hines, D. 1999,
ApJ, 512, 125

\bibitem[Schneider, Ehlers, \& Falco(1992)]{schneider92} Schneider, 
P., Ehlers, J., \& Falco, E.~E.\ 1992, Gravitational Lenses, Springer-Verlag
Berlin Heidelberg New York

\bibitem[Small, Sargent, \& Steidel(1997)]{small97} Small, 
T.~A., Sargent, W.~L.~W., \& Steidel, C.~C.\ 1997, \aj, 114, 2254 

\bibitem[Sprayberry et al.(1992)]{sprayberry92} Sprayberry, D. \& Foltz,
C. B. 1992, ApJ, 390, 39 

\bibitem[Tolea, Krolik, \& Tsvetanov(2002)]{tolea02}Tolea, A., Krolik,
J. H., \& Tsvetanov, Z. 2002, \apj, 578, L31  

\bibitem[Veron-Cetty \& Veron(2001)]{veron01} Veron-Cetty M.P., \&
Veron P. 2001, A\&A, 374, 92

\bibitem[Vignali et al.(2003)]{vignali03} Vignali, C., Brandt, W. N.,
\& Schneider, D. P. 2003, AJ, 125, 433 

\bibitem[Weymann, Morris, Foltz, \& Hewett(1991)]{weymann91} 
Weymann, R.~J., Morris, S.~L., Foltz, C.~B., \& Hewett, P.~C.\ 1991, \apj, 
373, 23 

\bibitem[Xanthopoulos et al.(1998)]{xanthopoulos98} Xanthopoulos, 
E.~et al.\ 1998, \mnras, 300, 649 

\bibitem[Yan, McCarthy, Storrie-Lombardi, \& Weymann(1998)]{yan98} Yan, L., 
McCarthy, P.~J., Storrie-Lombardi, L.~J., \& Weymann, R.~J.\ 1998, \apjl, 503,
L19

\bibitem[Zheng et al.(2000)]{zheng00} Zheng, W. et al. 2000, AJ, 120, 1607
\end{thebibliography}
\end{document}